# Thermal behavior of NIR active centers in Bi-doped optical fibers

Dvoretsky D.A.[1,3], Bufetov I.A.[1], Vel'miskin V.V.[1], Zlenko A.S.[1], Khopin V.F.[2],

Semenov S. L.[1], Guryanov A.N.[2], Denisov L.K.[3], Dianov E.M.[1].

*1- Fiber Optics Research Center of RAS, 38, Vavilov Str., Moscow, 119991, Russia; e-mail: iabuf@fo.gpi.ru,*

*2- Institute of Chemistry of High-Purity Substances of RAS; 49,Ttropinin Str. , N. Novgorod, 603600, Russia*

*3-Bauman Moscow State Technical University; 5, 2nd Baumanskaya Str., Moscow, 105005, Russia.*

**Abstract**

*The temperature dependences of optical loss and luminescence spectra have been measured in visible and NIR spectral range for Bi-doped silica and Bi-doped germanosilicate fibers for the first time. The temperature dependence of luminescence lifetime for Si-associated active bismuth centers in germanosilicate fiber was measured. It has been revealed, that distribution of $Bi^{3+}$ ions across the fiber preform is essentially different as compared to that of NIR active bismuth centers. Data received indicates that NIR-active centers are associated with the low oxidation state of bismuth ion and/or oxygen deficient center of the glass structure.*

# Оптические свойства волоконных световодов на основе плавленого кварца, легированного висмутом, в диапазоне температур от 300 К до 1500 К.

Д.А. Дворецкий[1,3], И.А. Буфетов[1], В.В. Вельмискин[1], А.С. Зленко[1], В.Ф. Хопин[2], С.Л. Семенов[1], А.Н. Гурьянов[2], Л.К. Денисов[3], Е.М. Дианов[1]

*1- Научный Центр Волоконной Оптики РАН, Россия, 119333, Москва, ул. Вавилова, 38; e-mail: iabuf@fo.gpi.ru*

*2- Институт Химии Высокохистых Веществ РАН, Россия, 603600, Н. Новгород, ул. Тропинина, 49;*

*3-Московский государственный технический университет имени Н.Э. Баумана, Россия, 105005, Москва, 2-я Бауманская ул., д.5*

*Аннотация*. Впервые измерены температурные зависимости полос поглощения и люминесценции в видимой и ближней ИК области спектра в легированных висмутом световодах из плавленого кварца и из германосиликатного стекла. Исследована зависимость от температуры времени жизни люминесценции ИК висмутового активного центра, ассоциированного с кремнием, в волоконном германосиликатном световоде. Обнаружено, что распределение ионов $Bi^{3+}$ по диаметру преформы световода из плавленого кварца значительно отличается от распределения ИК висмутовых активных центров, ассоциированных с кремнием. Полученные результаты указывают на то, что ион висмута в ИК активном центре должен быть в состоянии с *низкой валентностью и/или дефект сетки стекла, образующий вместе с ионом висмута ИК активный центр, является кислородно-дефицитным.*

## Введение.

Волоконные световоды на основе кварцевого стекла, легированного висмутом, являются оптическими широкополосными усиливающими средами [1]. Полоса длин волн, в которой наблюдается оптическое усиление в таких световодах, зависит от состава сердцевины световода, которая, как правило, состоит из алюмосиликатного, германосиликатного, фосфоросиликатного или германофосфоросиликатного стекла, легированного висмутом. И все эти световоды вместе в настоящее время обеспечивают оптическое усиление и лазерную генерацию в диапазоне от примерно 1.15 до 1.55 мкм. Кпд волоконных лазеров на ряде длин волн из указанного диапазона достигает 50%, а выходная мощность в непрерывном одномодовом режиме генерации составляет 20 Вт [2]. Недавно в висмутовом волоконном усилителе на германосиликатном световоде в полосе длин волн с шириной 40 нм и центральной длиной волны 1440 нм достигнуто усиление 25 дБ при мощности накачки всего 65 мВт [3]. Все это указывает на широкие возможности применения висмутовых световодов в качестве активных сред для усилителей и лазеров, работающих в новых частотных диапазонах как в непрерывном режиме, так и для генерации ультракоротких импульсов [4].

Но до настоящего времени ни одна из предложенных моделей ИК висмутовых активных центров (ВАЦ) в легированных висмутом световодах (или стеклах) не описывает адекватно имеющиеся экспериментальные данные [5, 6]. Данное обстоятельство является препятствием для дальнейшего совершенствования этого нового типа активных сред.

В последнее время были подробно исследованы люминесцентные свойства висмутовых световодов различного состава. Показано, что ВАЦ с наиболее простыми схемами энергетических уровней присутствуют в легированных висмутом световодах из чистого кварцевого стекла (SBi световоды), германатного (GBi световоды) и германосиликатного (GSBi световоды) стекла [7]. В алюмосиликатных и фосфоросиликатных световодах структура уровней существенно сложнее.

Исследование влияния температуры на свойства ВАЦ — один из путей, которые могут помочь в определении физической природы ВАЦ. Такие эксперименты для алюмосиликатных световодов для нескольких значений температур (77К, 300К, 600К и 1000К) были выполнены в [8]. Но их корректная интерпретация была затруднена ограниченным набором экспериментальных данных и сложной структурой уровней ВАЦ в стекле такого состава (по сравнению, например, с SBi световодами). Отметим, что изменение оптических потерь в волоконных световодах, не содержащих активных центров, при повышении температуры до 400 ºC исследовались ранее в работах, посвященных волоконным световодам с металлическими покрытиями [см., например, 9, 10]. При этом был обнаружен ряд интересных эффектов, влияющих на изменения оптических потерь в световодах в диапазоне до 100 дБ/км. В подобных экспериментах верхний предел температурного диапазона ограничен температурой плавления металлического покрытия.

В работах, посвященных разрушению световодов под действием лазерного излучения (так называемый fiber fuse effect) измерялись оптические потери в волоконных световодах (без ИК активных центров) при увеличении температуры примерно до температуры стеклования кварцевого стекла $T_g$ [11-13]. Измерения показали, что при нагревании до ≈1000 ºC оптические потери в кварцевых световодах не превышают величин порядка 10 дБ/км. Но при температуре около 1050 ºC [11] (или 1150 ºC, согласно [13]) начинается резкий рост потерь, которые уже через 50 ºC достигают величины 2000 дБ/км. Таким образом, при температурах до ≈1100 ºC и измерении оптических потерь на уровне, превышающем несколько дБ/м, можно использовать кварцевые волоконные световоды для подведения оптического излучения к исследуемому образцу световода, находящемуся в высокотемпературном объеме, без существенного искажения результатов измерений.

Пики поглощения ИК висмутовых активных центров в световодах имеют величины от $10^3$ до $10^6$ дБ/км (см.[7]), что указывает на желательность проведения измерений поглощения в висмутовых световодах именно в этом динамическом диапазоне. Для возможно более ясного проявления температурных зависимостей свойств ВАЦ необходимо проводить измерения до возможно более высоких температур, в случае стеклянных световодов характерным значением температуры в этой области является $T_g$.

Поэтому в настоящей работе были исследованы изменения спектров оптических потерь и люминесценции SBi и GSBi световодов от температуры в диапазоне от 20 до 1200°C.

**Экспериментальные образцы и условия проведения измерений.**

В качестве образцов для экспериментов были выбраны легированные Bi световоды из плавленого кварца без дополнительных добавок (2 образца), а также световод из плавленого кварца, солегированный германием (см.Таблица 1.). Причем эти световоды были получены по различным технологиям: преформа для световода 1 - по порошковой технологии (powder-in-tube, PIT) [14], а для световода 2 – по технологии FCVD (разновидность MCVD) [15]. Световедущие свойства сердцевины в световоде 1 обеспечивались за счет отражающей оболочки с пониженным коэффициентом преломления из кварцевого стекла, легированного фтором, а в световоде 2 – за счет формирования микроструктурированной оболочки с отверстиями, заполненными воздухом. Световод 3 изготавливался по технологии MCVD и его сердцевина на основе v-$SiO_2$ содержала 5 моль% германия, что обеспечивало формирование необходимого профиля показателя преломления в световоде. Внешний диаметр всех световодов составлял 125 мкм (без защитного полимерного покрытия). Отметим, что на всех трех типах световодов ранее была получена лазерная генерация в спектральном диапазоне около 1400-1500 нм [2, 14, 15].

Таблица 1. Обозначения, состав сердцевины и метод изготовления волоконных световодов

| № | Образец (наименование) | Состав сердцевины | Метод изготовления преформы | Ссылка |
|---|---|---|---|---|
| 1 | SBi-P | $100SiO_2-Bi_2O_3$ | PIT | [14] |
| 2 | SBi-H | $100SiO_2-Bi_2O_3$ | FCVD | [15] |
| 3 | GSBi | $95SiO_2-5GeO_2-Bi_2O_3$ | MCVD | [2,3] |

Состав сердцевины световода определялся с помощью сканирующего электронного микроскопа JSM 5910LV с рентгеновским спектроанализатором (Oxford Instruments). Содержание висмута во всех образцах было ниже порога регистрации (менее 0.02 ат.%), и поэтому его концентрация не отражена в Таблица 1.

В экспериментах для нагревания световодов использовалась трубчатая электрическая печь с рабочим объемом в виде цилиндра длиной 100 см и диаметром 2 см. Рабочий объем не был герметичен и всегда был заполнен лабораторным воздухом. Длина изотермической зоны, вдоль которой неоднородность температурного поля не превышала ±3ºC, составляла 40 см. Температура печи варьировалась в пределах от 25 ºC до 1200 ºC, средняя скорость нагрева в изотермической зоне в области температур 30 ºC – 700 ºC составляла 25 ºC/мин и плавно уменьшалась до 10 ºC/мин в области температур 700 ºC - 1200 ºC. Световоды сначала нагревались с указанными выше скоростями от комнатной температуры до 1200 ˚C или 1000 ˚C (в разных экспериментах), после чего медленно (в течение нескольких часов) охлаждались до комнатной температуры. Точность измерения температуры составляла ±0,5 ºC. Для проведения измерений исследуемый образец волоконного световода со снятым полимерным покрытием размещался по оси изотермической зоны в воздушной атмосфере. Его длина не превышала длины изотермической зоны. К концам световода приваривались отрезки неактивных световодов на основе плавленого кварца для ввода и вывода тестирующего излучения.

Спектры оптических потерь и люминесценции регистрировались в процессе нагрева световодов и после их охлаждения до комнатной температуры. Для этого в ИК области (750 – 1700 нм) использовался оптический анализатор спектра HP 70950B, а в области (350 – 850 нм) – анализатор спектра Ocean Optics S2000. Оптические потери в исходных образцах определялись стандартным способом (cut-back method, "метод облома"), основанном на сравнении спектров пропускания излучения через короткий и длинный отрезки световода. Ограничение длины исследуемого отрезка световода длиной изотермической зоны не позволяло измерять оптические потери в длинноволновом (850-1700 нм) диапазоне из-за сравнительно низкой их величины.

Спектры люминесценции висмутовых волоконных световодов регистрировались при различных температурах при накачке на длинах волн 457 нм и 808 нм в диапазонах от длины волны возбуждения до 1700 нм. В качестве источников возбуждения использовались излучение второй гармоники неодимового твердотельного лазера на длине волны 914 нм (мощность второй гармоники 100 мВт на длине волны 457 нм) и одномодовый полупроводниковый лазерный диод с волоконным выходом на длине волны 808 нм (выходная мощность 100 мВт). Для исключения влияния перепоглощения излучения люминесценции в сердцевине световода измерялся спектр излучения люминесценции, распространяющегося по кварцевой оболочке световода по схеме, аналогичной описанной в [16].

Для измерения зависимости времени релаксации люминесценции ИК активных висмутовых центров от температуры в диапазоне длин волн 1300-1500нм в качестве приемника излучения использовался InGaAs фотодиод с временем отклика 3 мкс. Для возбуждения люминесценции использовались прямоугольные импульсы с длительностью 10 мкс на длине волны 808нм. Сигнал регистрировался с помощью цифрового осциллографа (LeCroyWavepro 7100).

**Результаты измерений.**

На Рис. 1 показаны спектры оптических потерь рассматриваемых типов световодов, измеренные в видимой и ближней ИК области спектра при комнатной температуре. Спектры оптических потерь SBi-P и SBi-H световодов близки по форме и состоят из полос сложной формы с пиками на ~420нм, 820 нм и на ~1400нм на фоне монотонного повышения средних оптических потерь от 1600 нм до 400 нм примерно на 2 порядка. Световод GSBi отличается более низким средним уровнем потерь и несколько другой формой полос поглощения на ~420нм и ~1400нм. Что же касается полосы на 820 нм, то на длинноволновом ее крыле наблюдается дополнительно (по сравнению со световодами SBi-P и SBi-H) несколько слабо выраженных и менее интенсивных полос поглощения от 870 до ~1000 нм. Как было показано в [7], полосы поглощения в световоде GSBi на 460 нм и 940 нм связаны с ВАЦ, ассоциированными с атомами германия (Ge-ВАЦ), Люминесценция этих центров при комнатной температуре наблюдалась в области 950 нм и 1650 нм. Пики поглощения в световодах GSBi, SBi-P и SBi-H на ~420 нм, 820 нм и 1400 нм относятся к другим ВАЦ, связанным с кремнием (Si-ВАЦ). Si-ВАЦ люминесцируют в области 830 нм и 1400 нм. [7, 14]. Следует отметить, что на пики поглощения Si-ВАЦ в SBi-H и SBi-P на 1400 нм накладываются полосы поглощения групп OH с максимумом на 1380 нм. Положение упомянутых выше линий поглощения и люминесценции соответствует соответствует переходам между нижними четырьмя энергетическими уровнями Si-ВАЦ и Ge-ВАЦ, схемы расположения которых представлены на Рис. 1б. .В наших измерениях особое внимание было уделено исследованию изменения поглощения и люминесценции на длинах волн, соответствующих переходам в ВАЦ в SBi и GSBi световодах: в SBi наблюдались переходы, соответствующие только Si-ВАЦ, а в GSBi световодах – переходы, соответствующие как Si-ВАЦ, так и Ge-ВАЦ.

На рис. 2 показано изменение спектров оптических потерь в диапазоне 380нм-850нм в световодах из кварцевого стекла, легированных висмутом, SBi-H и SBi-P от температуры. Отметим, что в дырчатом световоде SBi-H каналы были заполнены воздухом при комнатной температуре и заварены (загерметизированы).

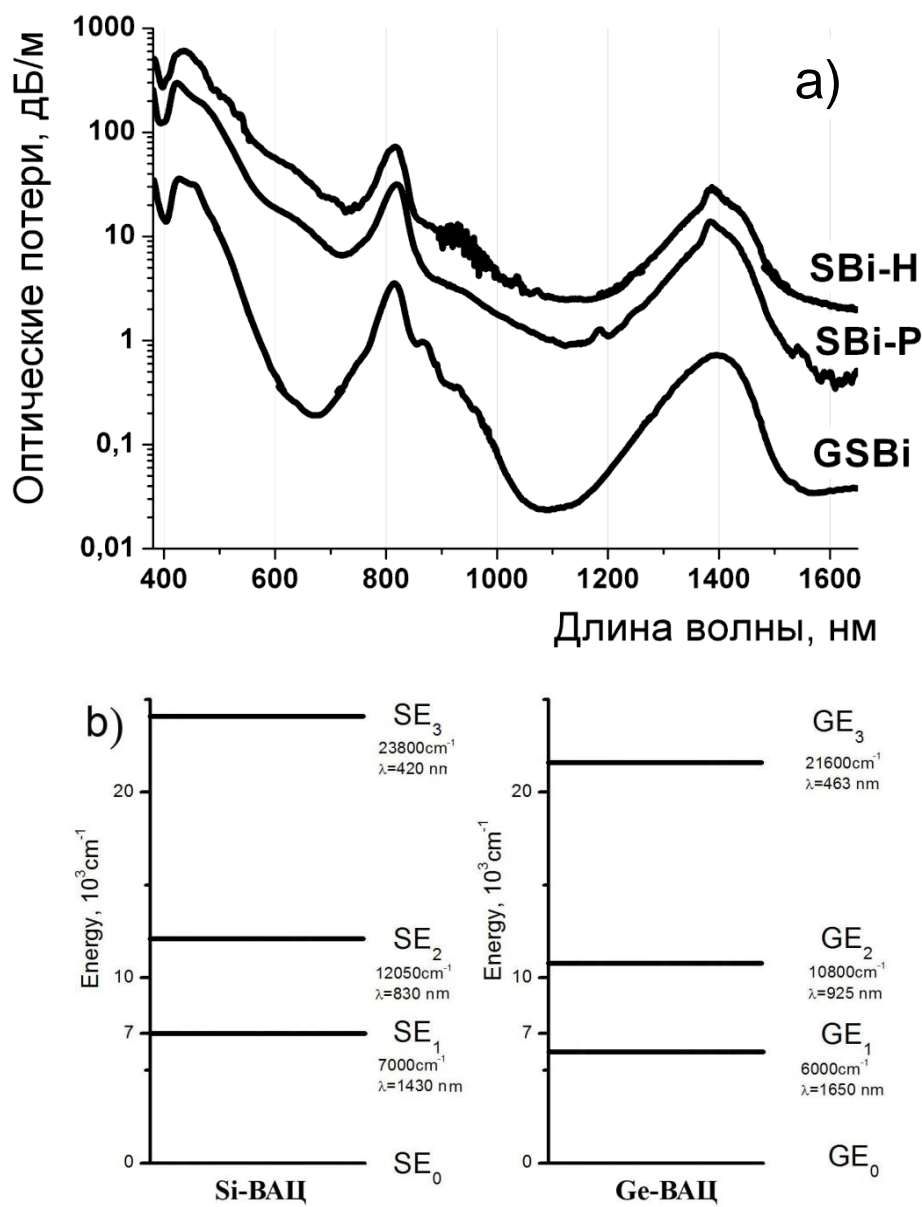

Рис. 1. a) Спектры оптических потерь германосиликатного световода GSBi и световодов на основе кварцевого стекла с добавкой висмута SBi-P и SBi-H. b) Схемы энергетических уровней Si-ВАЦ и Ge-ВАЦ (Справа приведены обозначения уровней, их расстояние от основного уровня и соответствующие длины волн излучения).

В световоде SBi-H (Рис. 2*а*) при нагреве наблюдается значительное уменьшение интенсивности полос поглощения в диапазоне длин волн 420 нм и 820 нм (переходы $SE_0 \rightarrow SE_3$ и $SE_0 \rightarrow SE_2$, см. Рис. 1б) после достижении температуры 700 ºС. Ограниченная длина световода в нагревателе не позволяла измерять оптические потери ниже 10 дБ/м. Одинаковая зависимость полос поглощения в области спектра ~ 420 нм и 820 нм от температуры (см. вставку на Рис. 2*а*) подтверждает принадлежность данных полос поглощения к одному ВАЦ. После охлаждения от 1200 ºC со скоростью 10 ºC/мин до комнатной температуры в SBi-H световоде поглощение на 420 нм и 820 нм составляло менее 10 дБ/м.

*а)* *б)*

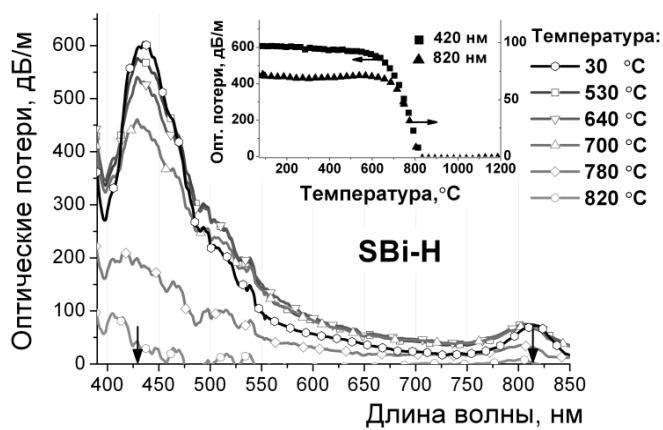 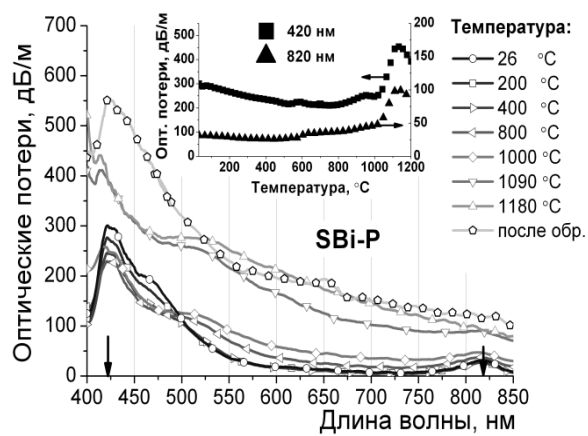

Рис. 2. Изменение спектра оптических потерь световодов при изменении температуры от 30 ºC до 1200 ºC. а)SBi-H; б)SBi-P. Вставки на рисунках показывают зависимости оптических потерь от температуры в полосах поглощения на указанных длинах волн.

В случае световода SBi-P (Рис. 2*б*) изменения поглощения в полосах на 420 нм и 820 нм значительно отличаются от аналогичных зависимостей для световода SBi-H. Здесь отсутствует резкое снижение оптических потерь в указанных полосах после 700ºC и наблюдается значительный их рост в области температур от ~ 1030 ºC до ~1130 ºC, выше которой они уменьшаются. После последующего охлаждения световода от температуры 1200ºC со скоростью 10 ºC/мин до комнатной наблюдалось увеличение уровня оптических потерь по сравнению со световодом до нагревания (см. рисунок 2*б* (после обр.)) практически во всем диапазоне измерений.

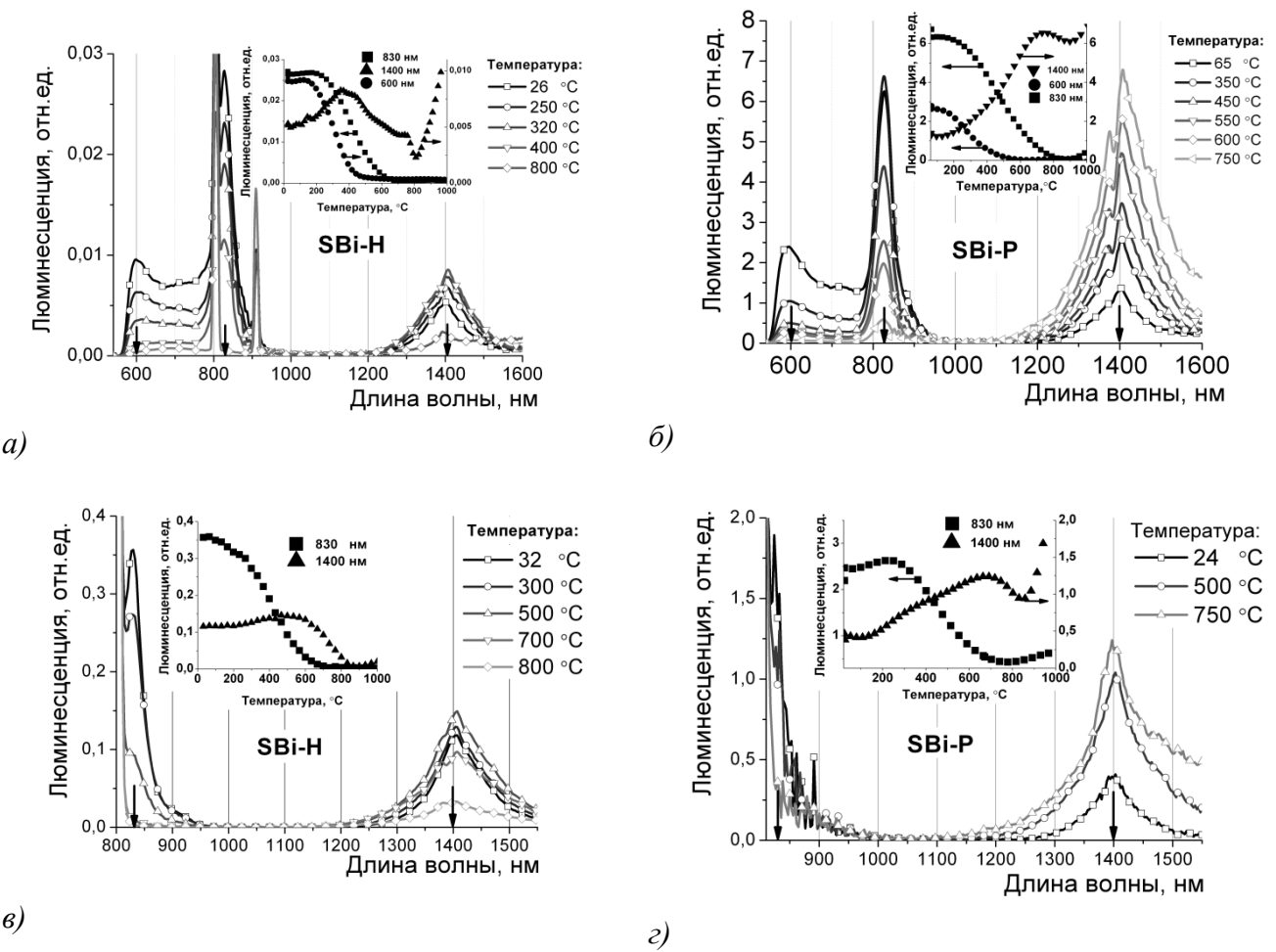

Рис. 3. Зависимость спектров люминесценции при возбуждении на 457 нм и 808 нм легированных висмутом световодов при изменении температуры от 30 °C до 1000 °C на воздухе: а – световод SBi-H при накачке на длине волны 457 нм; б – световод SBi-P при накачке на длине волны 457 нм; в – световод SBi-H при накачке на длине волны 808 нм; г – световод SBi-P при накачке на длине волны 808 нм.

На Рис. 3 показано изменение спектров люминесценции световодов при возбуждении на 457 нм и 808 нм при повышении температуры. На спектрах, кроме пиков люминесценции ВАЦ, наблюдаются полосы люминесценции на 600 нм, соответствующие люминесценции ионов $Bi^{2+}$ [7], а также узкие интенсивные пики на ~800 нм и ~915 нм, которые соответствуют рассеянному излучению накачки неодимового лазера и его первой гармоники. В световоде SBi-H при изменении температуры от 30 °C до 1000 °C интенсивность полосы люминесценции на 600 нм уменьшалась, начиная с температуры 200 °C, и снижалась до нуля (до порога регистрации) около 400 °C. Полоса люминесценции на 830 нм (переход $SE_2 \to SE_0$, см. Рис. 1б) плавно снижалась, начиная с температуры 300 °C, и достигала порога регистрации в районе температуры 800 ºC. Полоса люминесценции на 1400 нм (переход $SE_1 \rightarrow SE_0$) первоначально увеличивала интенсивность, достигала максимума около 400ºC, после чего плавно снижалась (до температуры ~740 ºC) и после этого резко снижалась до порога регистрации. Наблюдаемое нарастание уровня излучения на 1400 нм при температурах выше 820 ºC связано с нарастанием уровня теплового излучения (см. ниже). После охлаждения от температуры 1000 ºC со скоростью 10 ºC/мин до комнатной люминесценция в световоде SBi-H не наблюдалась.

На Рис. 3в показаны спектры люминесценции этого же световода при возбуждении на 808 нм при изменении температуры от 30 ºC до 1000 ºC. Спектр люминесценции в этом случае состоит из двух широких полос с максимумами на длинах волн 830 нм и 1400 нм, что соответствует данным, полученным ранее [7, 14]. Интенсивность люминесценции на 830 нм плавно снижалась при возрастании температуры от 30 ºC до 800 ºC, в то время как полоса люминесценции на 1400 нм сначала возрастала, достигала максимума около 500 ºC, и только затем плавно снижалась. И в этом случае после охлаждения световода от температуры 1000 ºC со скоростью 10 ºC/мин до комнатной люминесценция в световоде SBi-H также не наблюдалась.

На Рис. 3б,г показаны аналогичные зависимости спектров люминесценции для световода SBi-P. В этом световоде при накачке на длине волны 457 нм (Рис. 3б) полоса люминесценции на 600 нм также уменьшалась начиная с температуры 200 ºC и достигала порога регистрации около 400 ºC. Полоса люминесценции на 830 нм при повышении температуры, плавно уменьшалась и достигала порога регистрации около 800 ºC. Полоса люминесценции на 1400 нм, напротив, сначала значительно увеличивала свою интенсивность, достигала максимума при ~750 ºC, и после этого наблюдалось плавное ее снижение, а не резкое снижение до порога регистрации, как в случае световода SBi-H.

После охлаждения SBi-P световода от 1200ºC до комнатной температуры спектр его люминесценции практически не отличался от спектра люминесценции до нагревания, за исключением снижения интенсивности люминесценции полосы на 600 нм по сравнению с полосой на 830 нм.

Как показано выше, температура 700 ºC является характерным порогом, при котором значительно снижается поглощение в полосе 420 нм в случае световода SBi-H, а так же полоса люминесценции на 1400 нм при накачке на длине волны 457 нм быстро уменьшает свою интенсивность в районе температуры 700 ºC, что указывает на связь полосы поглощения на 420 нм с люминесценцией в полосе на 1400 нм, принадлежащих к единому ВАЦ.

Спектры люминесценции световода SBi-P при возбуждении на длине волны 808 нм показаны на рисунке 3г. Спектр люминесценции в этом случае состоит из двух широких полос с максимумами на длинах волн 830 нм и 1400 нм, которые изменяются с температурой практически так же, как и при возбуждении на 457 нм. Повышение интенсивности излучения на 1400 нм при приближении температуры к 1000ºC происходит из-за роста интенсивности теплового излучения.

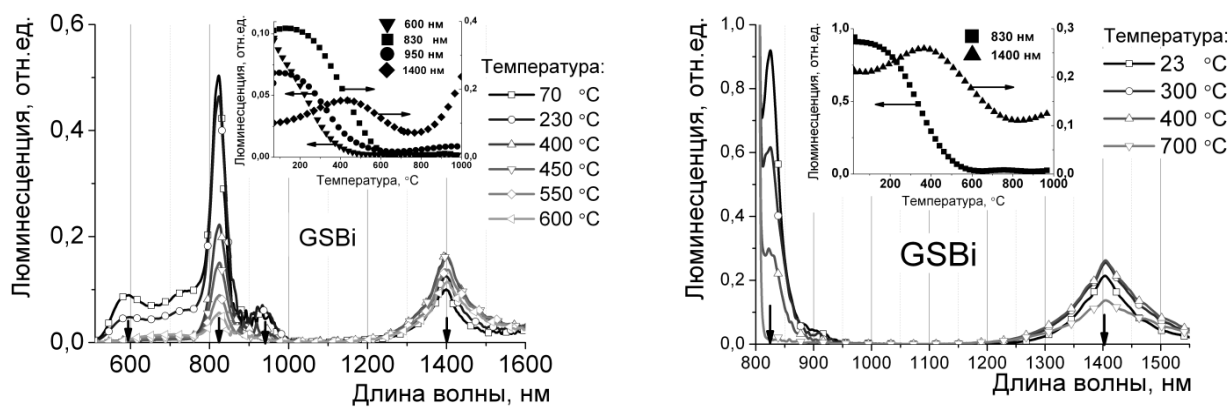

*а)*

*б)*

Рис. 4. Спектры люминесценции GSBiсветовода при ее возбуждении на длинах волн 457 нм и 808 нм при изменении температуры от 30 ºC до 1000 ºC на воздухе: а – при накачке на длине волны 457 нм; б–при накачке на длине волны 808 нм.

Результаты аналогичных экспериментов со световодом GSBi представлены на Рис. 4(а, б). По сравнению со световодами SBi-H и SBi-P, здесь наблюдались не только спектральные полосы люмнесценцииSi-ВАЦ и ионов $Bi^{2+}$ (на 600 нм), но и полоса люминесценции на 950 нм, принадлежащая Ge-ВАЦ. Поведение полос люминесценции Si-ВАЦ в этом световоде подобно поведению этих же полос в световоде SBi-P. Наиболее резким отличием является тот факт, что температурная зависимость полосы $Bi^{2+}$ на 600 нм имеет отрицательный наклон уже при комнатной температуре, в отличие от других исследованных световодов. Кроме того, максимум люминесценции на 1400 нм наблюдается при Т=450 °С (а не при Т=750°С, как в световоде SBi-P).

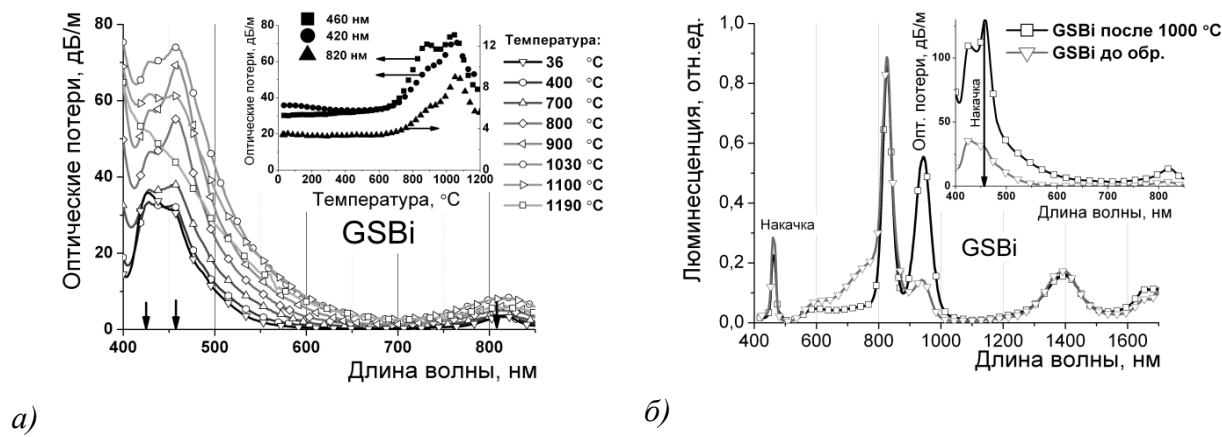

*а)*          *б)*

Рис. 5. а - Спектры оптических потерь световода GSBi для различных значений температуры световода, на врезке – зависимости оптических потерь световода GSBi от температуры для указанных длин волн.
б – спектры люминесценции световода GSBi до и после отжига при температуре 1000 °С (накачка на 457 нм); на врезке – спектры оптических потерь световода до и после отжига.

Зависимость оптических потерь в спектральном диапазоне длин волн от 400 нм до 850 нм в световоде GSBi от температуры 30 °С до 1200 °С представлена на Рис. 5*а*. Представляет интерес как абсолютное изменение оптического поглощения в полосах Si-ВАЦ (420 и 820 нм) и Ge-ВАЦ (460 нм), так и соотношение близких линий, а именно пары 420 нм – 460 нм. Поглощение на 460 нм возрастает быстрее, чем на 420 нм, начиная с ~700°С, после чего в полосе 460 нм наблюдаются два локальных максимума коэффициента поглощения: при 830°С и при 1030 °С. После 1030 °С во всех перечисленных линиях наблюдается снижение поглощения. В зависимостях поглощения от температуры для линий Si-ВАЦ (в отличие от линии Ge-ВАЦ на 460 нм) наблюдается только один максимум при ≈1030 °С.

На Рис. 5*б* представлен график люминесценции световода GSBi при комнатной температуре до и после процесса нагрева 1000 °С и последующего охлаждения. На вставке показаны спектры потерь световода GSBi до и после указанного цикла. Сравнение этих спектров с очевидностью показывает рост поглощения в полосе около 460 нм относительно полосы поглощения на 420 нм. В тоже время происходит увеличение интенсивности люминесценции в диапазоне длин волн около 950 нм при накачке на длине волны 457 нм.

Сравнение Рис. 5(а и б) показывает, что в световоде GSBi при нагревании происходит рост концентрации Si-ВАЦ, поглощающих в полосе на ~420 нм, и Ge-ВАЦ, поглощающих в полосе на ~460 нм [7], при этом люминесценция и поглощение Ge-ВАЦ относительно Si-ВАЦ увеличиваются в несколько раз.

**Обсуждение.**

При нагревании всех исследованных в настоящей работе световодов не наблюдается существенного изменения коэффициентов поглощения в линиях ВАЦ до температур ~700 °С. В то же время наблюдаются значительные изменения в интенсивности люминесценции на различных переходах ВАЦ. Так, во всех световодах наблюдается рост люминесценции на 1400 нм при одновременном снижении уровня люминесценции на 830 нм (см. Рис. 3, Рис. 4). Аналогичные результаты для кварцевого стекла, легированного висмутом по технологии SPCVD, были получены также в [17] в температурном диапазоне (20-600)°С. Для объяснения этого эффекта была исследована эволюция люминесценции в GSBi световоде на длине волны 1400 нм при накачке на 808 нм в зависимости от температуры. На Рис. 6 представлены полученные результаты. После импульса возбуждения длительностью 10 мкс наблюдается сначала нарастание люминесценции с коротким характерным временем (25 мкс и менее), зависящим от температуры, после чего люминесценция затухает с характерными временем ~600 мкс. Время разгорания люминесценции, которое в значительной мере характеризует время жизни Si-ВАЦ на уровне $SE_2$ (Рис. 1б), соответствующем поглощению на 820 нм и люминесценции на 830 нм, резко уменьшается с температурой и около 800°С становится менее 3 мкс. Данное обстоятельство объясняет снижение уровня люминесценции на 830 нм и одновременное увеличение интенсивности люминесценции на 1400 нм при нагревании: при увеличении температуры вероятность безызлучательной релаксации с уровня $SE_2$ на уровень $SE_1$ возрастает.

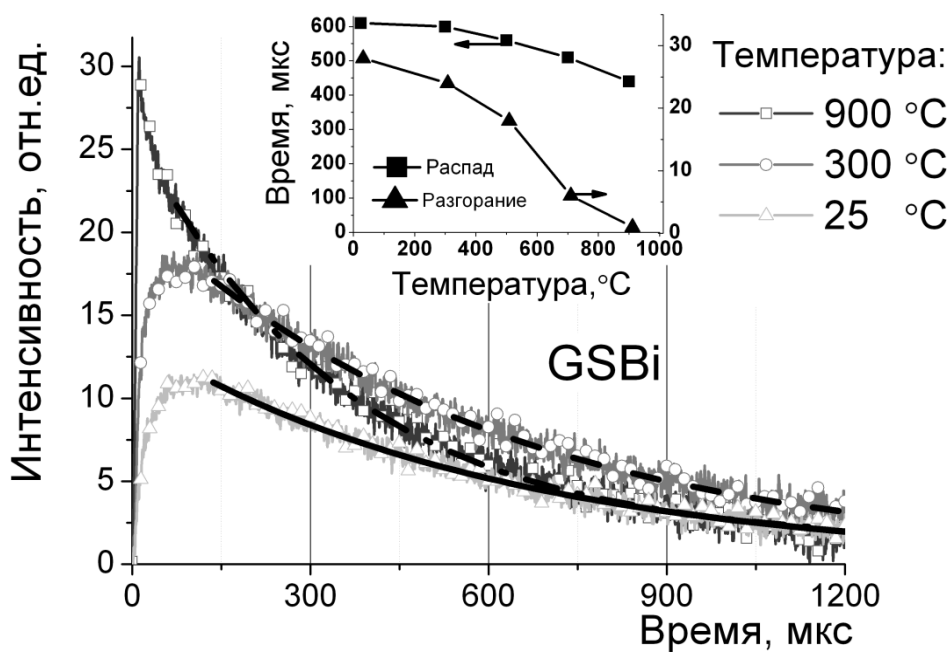

Рис. 6. Зависимость интенсивности люминесценции GSBi световода от времени при его возбуждении на длине волны 808 нм импульсом длительностью 10 мкс. На врезке: зависимость времени нарастания и релаксации люминесценции в полосе на 1400 нм от температуры при накачке на длине волны 808 нм.

В результате проведенных измерений было обнаружено снижение практически до нуля оптического поглощения и люминесценции на длинах волн Si-ВАЦ в световоде с каналами, заполненными воздухом и находящимися на расстоянии порядка 1 мкм от сердцевины световода SBi-H (диаметром около 6 мкм) при его нагревании до 800°C (Рис. 2а, Рис. 3а,в), что свидетельствует об исчезновении ВАЦ в сердцевине световода. В аналогичных условиях в световоде без каналов, заполненных воздухом (SBi-P), ВАЦ сохраняются (Рис. 2б, Рис. 3б,г). Отметим, что аналогичное явление было обнаружено ранее при вытяжке световода SBi-H в атмосфере кислорода [15]. Процесс вытяжки световода происходит при температуре около 2000 °C. Наши эксперименты показывают, что исчезновение ВАЦ имеет место при существенно более низкой температуре – около 800°C. Световоды SBi-H и SBi-P отличаются тем, что в первом расстояние от сердцевины до границы с газообразным кислородом составляет порядка 1 мкм, а во втором – около 60 мкм. Поэтому возникает предположение о том, что в SBi-H световоде диффузия кислорода при повышенной температуре приводит к разрушению ВАЦ. Правда, согласно [18], расстояние, на которое диффундирует кислород в v-$SiO_2$, при температуре 800°C за время ~15 мин., характерное для наших экспериментов, не превышает 0.03 мкм. Но в данных экспериментах диффузия происходит в материале световодов, структура стекла которых отличается от обычного v-$SiO_2$ большим количеством нарушенных связей и меньшей плотностью, так как получена в результате резкого охлаждения в процессе вытяжки [19]. В материалах, полученных подобным образом, коэффициент диффузии может быть существенно выше, чем в стекле, полученном в результате медленного охлаждения (см., например, [20]).

Таким образом, если принять в качестве исходной гипотезу о строении ВАЦ как сочетание иона висмута и дефекта сетки стекла [6], то полученные результаты указывают, что ион висмута в ВАЦ должен быть в состоянии с низкой валентностью (и его дополнительное окисление приводит к исчезновению центров), и/или связанный с ВАЦ дефект сетки стекла является кислородно-дефицитным.

Известно, что распределение кислородно-дефицитных центров в сердцевинах заготовок волоконных световодов неоднородно и имеет максимум на границе с отражающей оболочкой [21] (исследования выполнены для германосиликатных световодов). Поэтому мы провели дополнительные измерения распределения люминесценции на длине волны 830 нм при возбуждении на 457нм и распределения поглощения в максимуме на 230 нм по диаметру заготовки SBi-P световода.

Известно, что стекла, легированные висмутом, имеют полосу поглощения в УФ диапазоне, положение которой зависит от состава стекла [22] и находится обычно в диапазоне 200-300 нм. По-видимому, можно считать общепринятой точку зрения, согласно которой эта линия принадлежит иону $Bi^{3+}$ (см., например, [22, 23]). На вставке Рис. 7 показаны спектры поглощения сердцевины преформы световода SBi-P и, для сравнения, точно такой же преформы, но без легирования сердцевины висмутом. Очевидно, что полоса поглощения с центром на 230 нм связана с легированием Bi, и, согласно вышеизложенному, принадлежит иону $Bi^{3+}$. В таком случае зависимость коэффициента поглощения на длине волны 230 нм от поперечной координаты отображает распределение ионов $Bi^{3+}$ по поперечному сечению сердцевины преформы. В то же время, распределение яркости люминесценции на длине волны 830 нм должно (при однородном возбуждении) давать информацию о распределении Si-ВАЦ.

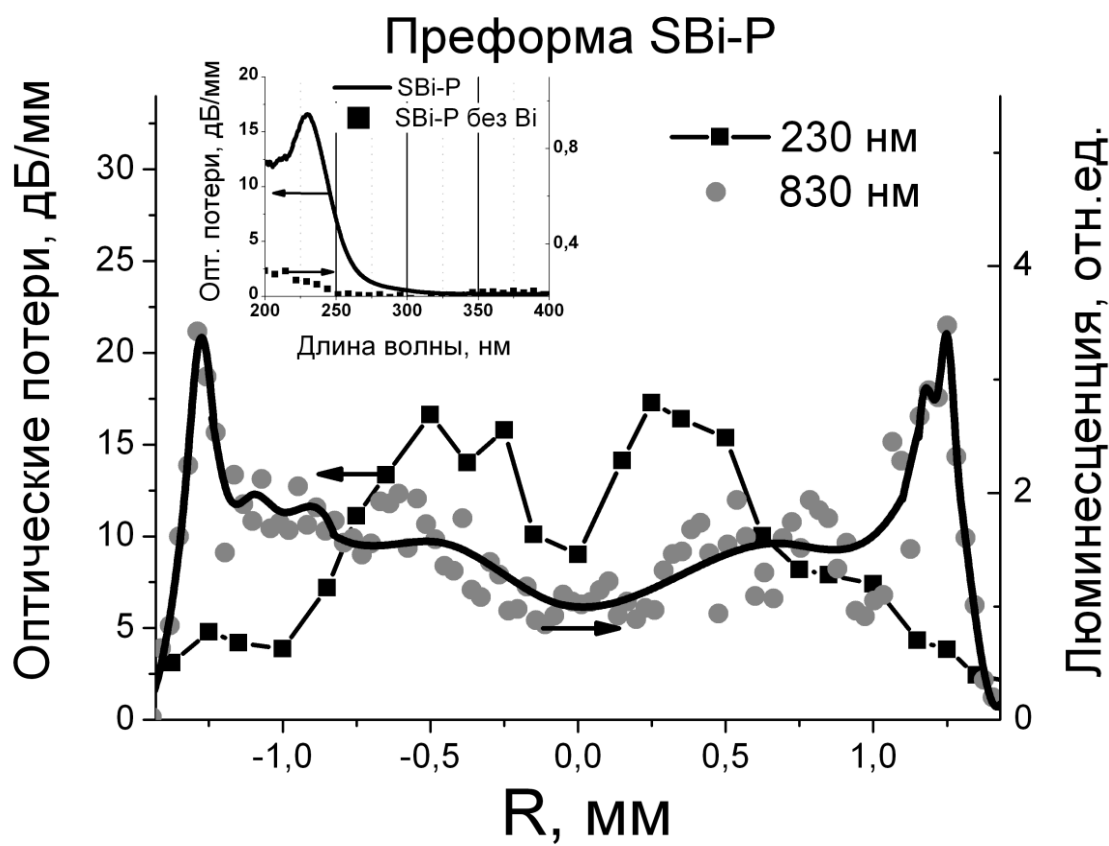

Рис. 7. Распределение поглощения в полосе на 230 нм и люминесценции в полосе на 830 нм по диаметру преформы световода SBi-P. На врезке показаны спектры оптических потерь в сердцевине преформы световода SBi-P и в точно такой же преформе, но без легирования висмутом.

На Рис. 7 представлены результаты измерений (пространственное разрешение в этих экспериментах составило ~0.1 мм). Полученные распределения существенно отличаются друг от друга. Если максимум концентрации $Bi^{3+}$ наблюдается в средней части сердцевины (с некоторым провалом в центре), то максимальная концентрация Si-ВАЦ (интенсивность распределения люминесценции на 830 нм) наблюдается в области границы сердцевины и отражающей оболочки. Причем повышенная концентрации Si-ВАЦ наблюдается в области толщиной ~0,4 мм от границы сердцевины. Таким образом, распределение ВАЦ по диаметру заготовки существенно отличается от распределения ионов $Bi^{3+}$, и достигает максимума вблизи границы с отражающей оболочкой, что качественно соответствует распределению кислородно-дефицитных центров в германо-силикатных заготовках [21]. Данное обстоятельство указывает на возможность участия кислородно-дефицитных центров стекла в формировании ИК висмутовых активных центров.

В настоящей работе наблюдается значительное температурное тушение люминесценции в полосе на 600 нм уже при 400 °C (см. Рис. 3 и Рис. 4), при этом различия зависимостей люминесценции на 600 нм и 830 нм от температуры как во время эксперимента в случае световодов SBi-P, SBi-H и GSBi, так и после термообработки при температуре 1200 °C световода SBi-P с последующим медленным охлаждением подтверждает принадлежность этих полос к различным активным центрам.

Обращают на себя внимание зависимости интенсивности люминесценции на 1400 нм при накачке на 457 нм для всех трех исследованных световодов. Для наглядности они приведены на одном графике Рис. 8. На вставке показана зависимость оптических потерь от температуры в полосах поглощения на 420 нм и 460 нм для световода GSBi. В случае световода SBi-H наличие воздуха вблизи сердцевины волоконного световода приводит к снижению полосы люминесценции на 1400 нм (Si-ВАЦ) после 400 °C. Отсутствие избыточного кислорода вблизи сердцевины волоконного световода в случае SBi-P приводит к снижению полосы люминесценции на 1400 нм только после ~750°C. Но аналогичная зависимость для световода GSBi близка по форме к зависимости для дырчатого световода SBi-H, хотя отверстия в световоде GSBi отсутствуют. По спектру поглощения (см. рис.8) видно, что в световоде GSBi концентрация Si-ВАЦ, поглощающих в полосе на 420 нм, снижалась при возрастании температуры до ~700 °C, а концентрация Ge-ВАЦ, поглощающих в полосе на 460 нм, увеличивалась. Таким образом, можно предположить, что $GeO_2$ в световоде GSBi служит источником кислорода для Si-ВАЦ, как и воздух в отверстиях световода SBi-P. Только в случае с GSBi световодом трансформация висмутовых активных центров происходит обратимым образом: после охлаждения световода ИК активные центры в нем сохраняются (см. рис.6б). Подобное поведение Si-ВАЦ и Ge-ВАЦ от температуры, с нашей точки зрения, также говорит в пользу модели висмутового активного центра, включающей ион висмута и кислородно-дефицитный дефект структуры стекла [6].

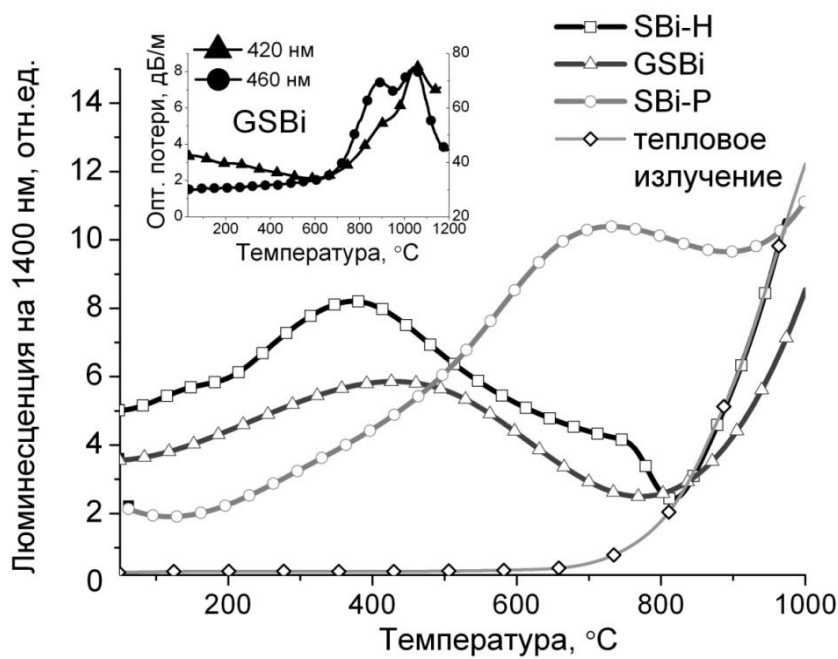

Рис. 8. Зависимость интенсивности люминесценции на 1400 нм от температуры при накачке на длине волны 457 нм для световодов SBi-H, SBi-P и GSBi и тепловой люминесценции для световода SBi-H. На вставке показана зависимость оптических потерь от температуры в полосах поглощения на 420 нм и 460 нм для световода GSBi.

На Рис. 8 для иллюстрации представлена также зависимость уровня теплового излучения на длине волны 1400 нм от температуры, полученная в тех же условиях, что и зависимость интенсивности люминесценции на 1400 нм световода SBi-H, но при выключенном источнике накачки. Сравнение зависимостей люминесценции световода SBi-H и теплового излучения показывает, что при температурах выше 800°C интенсивность теплового излучения существенно превосходит интенсивность люминесценции. Для других исследованных световодов и особенно для более коротких длин волн наблюдения относительный вклад теплового излучения составляет меньшую величину, однако наблюдается во всех экспериментах по регистрации люминесценции.

Здесь следует отметить тот факт, что в SBi-H световоде наблюдалась минимальная интенсивность люминесценции по сравнению с другими световодами, что увеличивало относительный вклад теплового излучения.

Представляет интерес эффект резкого увеличения поглощения в полосах 420 нм (Si-ВАЦ) и 460 нм (Ge-ВАЦ) световодов от температуры( начиная с Т=1030 ºС на врезке Рис. 2б. и с Т=700ºС на врезке Рис. 5а), и их снижение с температурой начиная с 1130 ºС в световоде SBi-P и с 1030 ºС в световоде GSBi. Возрастание поглощения при температурах, более низких, чем температура стеклования (отметим, что температура стеклования для чистого кварцевого стекла составляет ≈1202 ºС, а для чисто германатного≈513 ºС [24]) возможно как результат релаксации структуры стекла при медленном нагревании стекла после очень быстрого охлаждения, в данном случае в результате процесса вытяжки световода (см., например, [19, 20]). При приближении же к температуре стеклования, вследствие значительного нарушения структуры стекла наблюдается, по-видимому, снижение концентрации висмутовых активных центров.

**Выводы.**

Таким образом, в настоящей работе впервые измерены температурные зависимости полос поглощения и люминесценции в световодах из плавленого кварца, легированного висмутом, и в висмутовом германосиликатном световоде. Обнаружено значительное снижение полос поглощения и полос люминесценции висмутовых активных центров при нагревании дырчатого световода из плавленого кварца, легированного висмутом, в присутствии кислорода в диапазоне температур 700 – 800 ºС. При отсутствии доступа кислорода такой эффект не наблюдается. Данный результат вместе с результатами [15] указывают на то, что исследуемые ВАЦ связаны с ионами висмута в низком валентном состоянии ассоциированными с кислородно-дефицитными дефектами структуры стекла.

Обнаружено резкое снижение времени жизни Si-ВАЦ на уровне $SE_2$ (см. Рис. 1б) с 30 мкс до менее 3 мкс при повышении температуры от комнатной до 900 ºС. Время жизни люминесценции на 1400 нм (уровень $SE_1$) при этом снижается на 25%.

В германосиликатном световоде, легированном висмутом, при нагревании наблюдается рост поглощения во всех полосах активных висмутовых центров (как Si-ВАЦ, так и Ge-ВАЦ). При этом уровни поглощения и люминесценции Ge-ВАЦ относительно Si-ВАЦ увеличиваются в несколько раз.

Распределение ионов $Bi^{3+}$ по диаметру заготовки значительно отличается от распределения Si-ВАЦ. Обнаружено, что максимум концентрации Si-ВАЦ в световоде из плавленого кварца достигается у границы сердцевина-оболочка(аналогично концентрации ГКДЦ в германосиликатных световодах),что является еще одним указанием на возможную связь ВАЦ с кислородно-дефицитными дефектами структуры стекла.

# Л и т е р а т у р а

1 Bufetov I.A., Dianov E.M. Laser Physics Letters, **6**, 487 (2009).

2 Фирстов С. В., Шубин А. В., Хопин В. Ф., Мелькумов М. А., Буфетов И. А., Медведков О. И., Гурьянов А. Н., Дианов Е. М. Квантовая электроника, **41**, 581 (2011).

3 Melkumov M. A., Bufetov I. A., Shubin A. V., Firstov S. V., Khopin V. F., Guryanov A. N., Dianov . Optics Letters, **36**, 2408 (2011).

4 Dianov E.M., Krylov A.A., Dvoyrin V.V., Mashinsky V.M., Kryukov P.G. JOSA B, **24** 1807 (2007).

5 Peng M., Dong G., Wondraczek L., Zhang L., Zhang N., Qiu J. Journal of Non-Crystalline Solids, **357**, 2241 (2011).

6 Dianov E. M., Quantum Electron. **40(4)**, 283–285 (2010).

7 Firstov S. V., Khopin V. F., Bufetov I. A., Firstova E. G., Guryanov A. N., Dianov E. M. Optics Express, **19**, 19551 (2011).

8 Булатов Л.И., Машинский В.М., Двойрин В.В., Кустов Е.Ф., Дианов Е.М., Сухоруков А.П. Известия РАН, Серия физическая, **72**, 1751 (2008).

9 Bogatyrev V.A., Semenov S.L. В кн.: Spesialty Optical Fibers Handbook, Mendez A and Morse T.F., Editors, Elsevier, 798p., 2007

10 Voloshin V.V., Vorob'ev I.L., Ivanov G.A., Isaev V.A., Kolosovskii A.O., Lenardich B., Popov S.M., ChamorovskiiYu.K. Journal of Communications Technology and Electronics, **56**, 90 (2011).

11 Kashyap P. Proc. International Conference Lasers'87, (Lake Tahoe, Nevada, Dec. 7-11, 1987), p.859 (1987).

12 Hand D.P., Russell P.St.J., Optics Letters, **13**, 767 (1988).

13 Davis D.D., Meller S.C., DiGiovanni D.J. Proc. SPIE, **2966**, 592 (1997).

14 Bufetov I.A., Melkumov M.A., Firstov S.V., Shubin A.V., Semenov S. L., Vel'miskin V.V., Levchenko A.E., Firstova E.G., and Dianov E. M. Optics Letts, **36**,166 (2011)

15 Zlenko A.S., Dvoyrin V.V., Mashinsky V.M. et al. Optics Letts, **36**,2599 (2011)

16 Mattsson K.E. Optics Express, **19**, 19797 (2011).

17 Базакуца А.П., Бутов О.В., Голант К.М. Третья Всероссийская конференция по волоконной оптике, 12-14 октября 2011, Пермь, опубликовано Фотон-Экспресс, №6, 108 (2011).

18 Williams E.L. Journal of the American Ceramic Society, **48**, 190 (1965).

19 Vogel W. Glass chemistry. Springer-Verlag, Berlin Heidelberg, 464p. (1994).

20 Mehrer H. Diffusion in solids. Springer-Verlag, Berlin, Heidelberg (2007).

21 Neustruev V.B. J.Phys.: Matter, **6**,6901 (1994)

22 Duffy J.A., Ingram M.D. Journal of chemical physics, **52**, 3752 (1970)

23 Denker B.I., Galagan B.I., Shulman I.I., Sverchkov S.E., Dianov E.M. Applied Physics B, **103**, 681 (2011).

24 Ozhovan M.I. Journal of Experimental and Theoretical Physics, **103**, 819 (2006).